\documentclass[preprint,showpacs,preprintnumbers,amsmath,amssymb]{revtex4}
\usepackage{latexsym,amsmath,graphics,graphicx}

\newcommand{\loc}{\mathrm{local}}
\newcommand{\glob}{\mathrm{global}}
\newcommand{\bs}{\boldsymbol}

\begin{document}
\title{Mapping the magnetic exchange interactions from first principles: 
       Anisotropy anomaly and application to Fe, Ni, and Co}
\author{Samir Lounis$^1$}\email{slounis@uci.edu}
\author{Peter H.~Dederichs$^2$}
\affiliation{$^1$Department of Physics and Astronomy, University of 
California Irvine, California 92697, USA}
\affiliation{$^2$Institut f\"ur
Festk\"orperforschung, Forschungszentrum J\"ulich, D-52425 J\"ulich,
Germany}


\begin{abstract}
Mapping the magnetic exchange interactions from model Hamiltonian to density functional 
theory is a crucial step in multi-scale modeling calculations. Considering the usual 
magnetic force theorem but with arbitrary rotational angles of the spin moments, a spurious 
anisotropy in the standard mapping procedure is shown to occur 
provided by bilinear-like contributions of high order spin interactions. The evaluation of 
this anisotropy gives a hint on the strength of non-bilinear terms characterizing the 
system under investigation. 

\end{abstract}

\maketitle

Multiscale modeling approaches are extremely important for describing 
huge magnetic systems, e.g. at the micrometer-scale which would be impossible with 
only density functional theory (DFT). In magnetism, usually the multiscale approach is 
performed after mapping the magnetic exchange 
interactions (MEI) of a classical Heisenberg model to the DFT counterparts. This is 
a crucial task which can lead to wrong results if not done carefully. The simple model 
is described by 
\begin{equation}\mathcal{H} = - \, \frac{1}{2} \ 
\sum_{i\not= j} \ J_{ij} {\vec e}_i \cdot {\vec
e}_j,\label{heisenberg-model} \end{equation} 
where $J_{ij}$ describes the pairwise (two-spin) MEI between 
spins at lattice sites $i$ and $j$ while ${\vec e}_i (1,\theta,\phi)$ defines 
the direction of the local moment $\vec{M_i}$. 
Sometimes, higher order terms such as the four-spin or the biquadratic MEI 
are introduced in the previous Hamiltonian for a 
better mapping of the DFT results\cite{kurz,fahnle}.

Once the MEI extracted, the investigation of magnetism of several type of 
systems can be performed going from molecules\cite{molecule}, 
transition metals alloys\cite{alloys,alloys2} and 
surfaces\cite{surface,surface2}, diluted magnetic semiconductors\cite{dms1,dms2}, to 
clusters\cite{cluster1,cluster2,cluster3,cluster4} and even for strongly correlated stytems\cite{savrasov2}. 
Thermodynamical properties are then easily accessible  
such as Curie temperatures, specific 
heat or magnetic excitation spectra and spin waves stiffness 
in multi-dimensional systems.

An elegant method to extract the MEI is based on a
 Green function technique which 
has been derived 20 years ago by Lichtenstein and 
coworkers\cite{lichtenstein} 
(noted in the text LKAG). 
Instead of calculating several magnetic configurations, this method, based on 
the magnetic force theorem (MFT)\cite{heine,oswald}, 
allows the evaluation of the MEI from one collinear configuration which is usually 
ferromagnetic. Computationally, this method is thus very attractive. 

Assuming infinitesimal rotation angles of the magnetic moments 
(limit of infinite magnon wavelength) is necessary 
to get the final LKAG formula for the MEI. 
However, one should note that this formalism is used for arbitrary big rotation angles 
(finite magnon wavelength) as 
well. Thus, many improvements of the formalism have been proposed recently: 
Bruno\cite{bruno} proposed a renormalized MFT using the constrained DFT\cite{dederichs} 
leading to unrealistic high LDA Curie temperature ($T_c$) for fcc Ni. The same effect has been 
observed using the proposal of Antropov\cite{antropov2003}. 
Katsnelson and Lichtenstein proposed in their recent 
publication\cite{katsnelson} a reconciliation between the old 
formalism\cite{lichtenstein} and the new renormalized 
theories\cite{bruno,antropov2003}. 
They have shown that the improvements proposed are well suited for the static 
response function while the LKAG formalism is optimal for 
calculations of the magnon spectra. A more rigorous approach 
is based on the calculation of 
dynamical transverse susceptibility\cite{callaway,savrasov,mills,staunton,buczek,
sasioglu,lounis_TDDFT} which 
is computationally more involved.

In the present contribution, we revisit the LKAG formalism and scrutinize 
one of the first assumptions assumed in the mapping procedure which has not 
been discussed yet. We demonstrate that an 
interesting issue occurs  
in the original mapping and thus in 
the majority of improvements as well. Avoiding the long 
wave or the infinitesimal rotation angle approximation, 
an anisotropy of the 
DFT MEI is obtained. This inconsistency is interpreted as a contribution to the DFT mapped part 
from high order MEI, such as 
the four-spin interactions, but behaving like bilinear terms.

In our demonstration we follow the usual mapping procedure with three steps 
to consider: (i) definition of the 
classical Heisenberg model, (ii) evaluation of the DFT counterpart, (iii) 
mapping and extraction of the MEI.

{\bf Classical Heisenberg model for pair interactions}. As done in LKAG, we consider 
eq.~\ref{heisenberg-model} and determine 
the rotation energy of two spin moments at sites $i$ and $j$, 
which are initially ferromagnetically aligned.  
Contrary to LKAG, here we assume different rotation angles for $i$ and $j$. 
First, we determine the energy difference 
between this new magnetic state and the ferromagnetic one 
\begin{eqnarray}
\Delta E_{i+j}&=&
-\sum_{\substack{n\not=i\\ n\not=j}} J_{in}(e_i^z-1)
-\sum_{\substack{m\not=i\\ m\not=j}} J_{mj}(e_j^z-1) \nonumber\\
&&-J_{ij}(\vec{e}_i\cdot \vec{e}_j-1)
\end{eqnarray}
where the $z$-axis refers to the quantization axis of the ferromagnetic environment and $n$ and $m$ 
to environmental atoms.
Second, since we are interested in the MEI between atom $i$ and atom $j$ 
we subtract the interaction energies 
($\Delta E_i$ and $\Delta E_j$) of each atom with the 
environment. This is obtained after 
rotating only one of the two atoms, by the same angle as assumed 
for $\Delta E_{i+j}$.
\begin{eqnarray}
\Delta E_i &=& -\sum_{\substack{n\not=i\\n\not=j}} 
J_{in}(e_i^z-1) - J_{ij}(e_i^z-1)\label{energy-diff-basic}
\end{eqnarray}

The final quantity which depends only on the MEI is thus given by
\begin{eqnarray}
\Delta E_{(i,j)}&=&\Delta E_{i+j}-
\Delta E_i-\Delta E_j\label{eq:2}\\
\Delta E_{(i,j)}&=&-J_{ij} \bigg[ 1 +  \cos(\theta_i)\cos(\theta_j) - 
\cos(\theta_i) -\cos(\theta_j) \nonumber \\
&&+ \sin(\theta_i) \sin(\theta_j) 
\cos(\phi_i-\phi_j)\bigg ]\label{eq-heisenberg}
\end{eqnarray}
if polar and azimuthal angles ($\theta_i,\phi_i$) and ($\theta_j,\phi_j$) are 
introduced. In their work,  LKAG cant the two 
spins by an equal angle $\theta$ but in opposite directions {\it i.e.} by setting 
$\theta_i=\theta_j=\theta$ when 
evaluating $\Delta E_{i}$ and $\Delta E_{j}$ while they cant the two 
spins by $\theta/2$ and consider $\phi_i-\phi_j=\pi$ when evaluating $\Delta E_{i+j}$. One then obtains 
$\Delta E_{(i,j)} = -J_{ij}\bigg[ 1 -\cos(\theta)\bigg ]$ in agreement with LKAG. (Note that in the DFT counterpart expression LKAG use an angle 
$\theta/2$ for $\Delta E_i$ and $\Delta E_j$ instead of $\theta$). For small rotations $\theta_i$, $\theta_j$  
eq.~\ref{eq-heisenberg} simplifies to
\begin{eqnarray}
\Delta E_{(i,j)}&\approx& -J_{ij}  
\theta_i \theta_j \cos(\phi_i-\phi_j)
\end{eqnarray}

{\bf Magnetic pair interaction from DFT.} This difference is directly 
given by
\begin{eqnarray}
\Delta E_{(i,j)}&=&\int^{E_F} dE (E-E_F) \Delta n_{(i,j)}(E) \nonumber \\ 
&=&-\int^{E_F}dE \Delta N_{(i,j)}(E),
\end{eqnarray}
with $\Delta N_{(i,j)}(E)$ being the corresponding change 
of the integrated density of states (IDOS) and $E_F$ being the Fermi energy.
\begin{equation}
\Delta N_{(i,j)}(E)=\Delta N_{i+j}(E)- \Delta N_{i}(E)- \Delta N_{j}(E),
\label{integrated-density}
\end{equation}
Hence, $\Delta N_{i+j}(E)$ is the change of the 
IDOS when both atoms $i$ and $j$ have their moments 
rotated. 
$\Delta N_{i}(E)$ and $\Delta N_{j}(E)$ 
are changes of the IDOS when only one moment is rotated. 
$\Delta N_{(i,j)}(E)$ is the change of the IDOS corresponding 
to the interaction energy between the moments $i$ and $j$ as expressed in 
eq.~\ref{eq-heisenberg}.

Now, we can calculate every term in eq.~\ref{integrated-density} using multiple 
scattering theory and take advantage of the Lloyd's formula\cite{lloyd,drittler}:
\begin{equation}
\Delta N(E)=-\frac{1}{\pi}\mathrm{Im}\ \mathrm{Tr_{nLs}}\ 
\mathrm{ln}\ (\bs{1}-\bs{{G}}(E) \Delta {\bs V}),
\end{equation}
where the trace Tr is taken over the site (n), orbital momentum (L) and spin (s) indices. 
Knowing the Green function $\bs{G}$ of the 
initial system describing the 
collinear magnetic state, this formula allows an exact determination of the 
change in the IDOS just by knowing the potential difference $\Delta {\bs V}$ 
induced by the rotation of a magnetic moment.

When rotating the magnetic moments of two atoms $i$ and $j$, the 
interactive part of the 
integrated density of states 
according to eq.~(\ref{integrated-density}) is given 
by
\begin{eqnarray}
\Delta N_{(i,j)}(E)&=&-\frac{1}{\pi} \mathrm{Im} \ \mathrm{Tr_{nLs}} \bigg [ 
\mathrm{ln}
\bigg{(}
\bs{1}-\bs{{G}}(E) (\Delta {\bs V_i}+ \Delta {\bs V_j})\bigg{)}\nonumber \\ 
&&-\mathrm{ln}
\bigg{(}(\bs{1}-\bs{{G}}(E) \Delta {\bs V_i})(\bs{1}-\bs{{G}}(E) \Delta {\bs V_j})\bigg{)} 
\bigg ]\nonumber
\end{eqnarray}

After taking the trace over n, the formulation giving the IDOS can be simplified into:
\begin{eqnarray}
\Delta N_{(i,j)}&=&
-\frac{1}{\pi} \mathrm{Im} \ \mathrm{Tr_{Ls}} \ \mathrm{ln}
\bigg{(}
\bs{1}-\frac{\Delta\bs{t}_i\bs{{G}}_{ij}\Delta\bs{t}_j\bs{{G}}_{ji}}
{(\bs{1}-\Delta\bs{t}_i\bs{{G}}_{ii})(\bs{1}-\Delta\bs{t}_j\bs{{G}}_{jj})}
\bigg{)}\label{integrated-density-2},
\end{eqnarray}
which is equivalent to eq B.1 from LKAG. Here we dropped out the 
argument E for reasons of clarity and 
the scattering t-matrices $\Delta \bs{t}_i$ and $\Delta \bs{t}_j$ 
describe all scattering 
processes at the isolated atoms $i$ and $j$. $\Delta \bs{t}$ is defined by 
$\Delta \bs{V}/(1-\bs{G}\Delta \bs{V})$.

The term $\bs{{G}}_{ji}\Delta\bs{t}_i\bs{{G}}_{ij}\Delta\bs{t}_j$ describes the 
scattering of an electron at a site j, the propagation to the site i from which it 
is scattered back to site j. It is a second order process which is expected to be 
very small compared to 1. A similar argument can be used for the denominator. Indeed, if one makes a 
Taylor expansion of the denominator, terms like 
$\bs{{G}}\Delta\bs{t}\bs{{G}}\Delta\bs{t}\bs{{G}}\Delta\bs{t}$ would appear but are third order processes 
and thus 
are expected to be much smaller than 1. 

After a first order expansion of 
eq.~\ref{integrated-density-2} we obtain
\begin{equation}
\Delta N_{(i,j)}\sim\frac{1}{\pi} \mathrm{Im} \ \mathrm{Tr}_{Ls} \
\bs{{G}}_{ji}\Delta\bs{t}_i\bs{{G}}_{ij}\Delta\bs{t}_j
\label{integrated-density-3}
\end{equation}

The previous equation is expressed in the global spin 
frame of reference, {\it i.e.}
the t-matrices have non-diagonal elements which is not the 
case of the magnetically 
collinear host Green function $\bs{{G}}$. The MFT states that the spin-moment 
does not change upon rotation, meaning that the t-matrix within the local 
spin frame of reference of each atom does not change. Once calculated in the 
initial collinear state, the t-matrix is easily obtained:
\begin{eqnarray}
\bs{t}_n^{\glob}(E)&=&\frac{1}{2}\left[t^{\loc}_{sum}(E) \bs{1} +
t^{\loc}_{diff}(E)\bs{U}_n \bs{\sigma}_z \bs{U}_n^{\dag} \right],
\label{eq:10-bis2}
\end{eqnarray}
with $\bs{U}$ being a rotation matrix defined as following 
\begin{eqnarray}
\bs{U} &=& \begin{bmatrix}
{\cos(\frac{\theta}{2}) e^{-\frac{i}{2}\phi}}&
{-\sin(\frac{\theta}{2}) e^{-\frac{i}{2}\phi}} \\
{\sin(\frac{\theta}{2}) e^{\frac{i}{2}\phi}}&
{\cos(\frac{\theta}{2}) e^{\frac{i}{2}\phi}} 
\end{bmatrix} .
\label{eq:9}
\end{eqnarray}
 and $t^{\loc}_{sum}$ and 
$t^{\loc}_{diff}$ are equal to respectively 
$t^{\loc}_{\uparrow}+t^{\loc}_{\downarrow}$ and 
$t^{\loc}_{\uparrow}-t^{\loc}_{\downarrow}$. From the new t-matrix we subtract 
the initial one needed in eq.~\ref{integrated-density-3} 
\begin{eqnarray}
\Delta \bs{t}_i^{\glob}(E)&=&\frac{1}{2}\Delta t_{diff}^i(E)
\begin{bmatrix}
\cos(\theta_i) -1 & \sin(\theta_i) e^{-i{\phi_i}}\\
\sin(\theta_i) e^{i{\phi_i}} & -\cos(\theta_i)+1
\end{bmatrix}\label{tmatrix1},
\end{eqnarray}
which is inserted in eq.~{\ref{integrated-density-3}} leading to 
\begin{eqnarray}
\Delta N_{(i,j)}&\sim&\frac{1}{4\pi} \mathrm{Im} \ \mathrm{Tr_{L}}\
\bigg [(\bs{A}+\bs{C})(\cos(\theta_i)-1)(\cos(\theta_j)-1) \nonumber \\
&&+
2\bs{B}\sin(\theta_i)\sin(\theta_j)\cos(\phi_i-\phi_j)\bigg ]
\end{eqnarray}
after taking the trace over the spins with 
\begin{eqnarray}
\bs{A}&=&\bs{{G}}_{\uparrow}^{ij}\Delta t_{diff}^j
\bs{{G}}_{\uparrow}^{ji}\Delta t_{diff}^i,\  
\bs{B}=\bs{{G}}_{\uparrow}^{ij}\Delta t_{diff}^j
\bs{{G}}_{\downarrow}^{ji}\Delta t_{diff}^i, \nonumber \\  
\bs{C}&=&\bs{{G}}_{\downarrow}^{ij}\Delta t_{diff}^j
\bs{{G}}_{\downarrow}^{ji}\Delta t_{diff}^i.
\end{eqnarray}

{\bf Mapping.} Thus, the energy difference is given by
\begin{eqnarray}
\Delta E_{(i,j)}&=&-J_1 (1+\cos(\theta_i)\cos(\theta_j)-\cos(\theta_i)-\cos(\theta_j)) \nonumber \\
&& - J_2 \sin(\theta_i)\sin(\theta_j)\cos(\phi_i-\phi_j),
\label{last-lloyds}
\end{eqnarray}
where $J_1 =\frac{1}{4\pi} \mathrm{Im} \ \mathrm{Tr_{L}}\ \int^{E_F}dE (\bs{A}+\bs{C})$ 
and $J_2 = \frac{1}{4\pi} \mathrm{Im} \ \mathrm{Tr_{L}}\ \int^{E_F}dE  2\bs{B}$. 
This DFT expression is incompatible with expression (\ref{eq-heisenberg}) calculated from the Heisenberg model, since two parameters $J_1$ and $J_2$ appear. Note that LKAG give only the expression for $J_2$, which is also the expression used in the literature. However, it is only the correct expression 
for small angles $\theta_i$, $\theta_j$, since $J_1$ varies as 
$\theta_i^2\theta_j^2$. We face here an important dilemma in determining the MEI, which, as we will show, results from higher spin interactions automatically included in the second order DFT approach. 

Let us evaluate the difference between the 
two terms:
\begin{eqnarray}
J_1-J_2&=& \frac{1}{4\pi} \mathrm{Im} \ \mathrm{Tr_{L}}\ \int^{E_F}dE \nonumber \\
&&
({{G}}_{\uparrow}^{ij}-{{G}}_{\downarrow}^{ij})
\Delta t_{diff}^j
({{G}}_{\uparrow}^{ji}-{{G}}_{\downarrow}^{ji})
\Delta t_{diff}^i
\end{eqnarray}
Since agreement with the Heisenberg model is only obtained, if $J_2=J_1$ 
or $A+C=2B$, the difference $J_1-J_2$ vanishes only if 
${{G}}_{\uparrow}={{G}}_{\downarrow}$ {\it i.e.}
for a non-magnetic reference system. This means that any 
magnetic system would lead to two possible values for 
the MEI. It is true that for magnetic excitations with tiny rotation angles or 
for what is called the long wavelength approximation (LWA), 
one gets rid off the first term 
in eq.~\ref{last-lloyds} but the error grows like 
$(J_1-J_2) (\cos(\theta_i)-1)(\cos(\theta_j)-1)$. If the 
desired excited magnetic state is close to high values of the 
rotation angle then both terms $J_1$ and $J_2$ 
have to be considered.

\begin{figure}
\begin{center}
\includegraphics*[width=1.\linewidth]{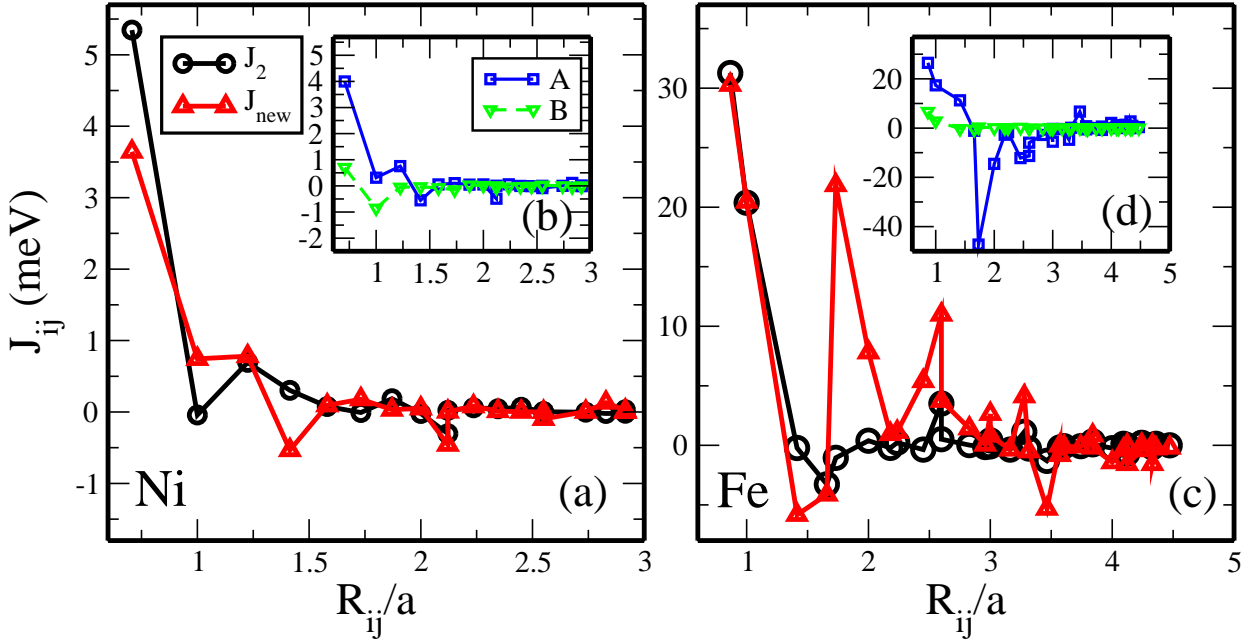}
\end{center}
\caption{The MEI $J_2$ (circles) and $J_{\mathrm{new}}=J_2+\frac{J_2-J_1}{2}$  (triangles)  
 calculated for bulk fcc Ni (a) and bcc Fe (c) with respect to the distance $R$ with $a$ being 
the lattice parameter. In the insets are plotted the terms $A$ and $B$ 
for Ni (b) and Fe (d). For reasons given below, the 
discrepancy between $J_2$ and $J_{\mathrm{new}}$ is much stronger for Fe compared to Ni.}
\label{jij_bulk}
\end{figure}

Using the full-potential Korringa-Kohn-Rostoker Green 
function method\cite{SKKR} within the local density approximation (LDA)\cite{LDA} or the generalized 
gradient approximation (GGA)\cite{GGA}, we evaluated these terms for 
usual bulk systems: Ni and Fe (see Fig.~\ref{jij_bulk}) 
and found that $J_1$ and $J_2$ are, 
on one hand,  
relatively similar for Ni since it has very small magnetic 
moments (0.61 $\mu_B$).  On the other hand, Fe bulk 
is characterized by a stronger discrepancy due to its high 
bulk magnetic moments (2.3 $\mu_B$).

In order to grasp some insight on the first term $J_1$ we propose to consider from 
the model Hamiltonian side terms beyond the Heisenberg model which are expected to be implicitly included in 
the DFT counterpart. The additional terms can be obtained 
from a perturbation expansion of the Hubbard model\cite{takahashi,kurz}. The 
first terms which have been added are 
the four-spin interactions, $H_{\mathrm{4-spin}}=-
\sum_{\substack{m\not = n\not = p\not = q}}K_{mnpq}
[(\vec{e}_m\vec{e}_n)(\vec{e}_p\vec{e}_q)+
(\vec{e}_n\vec{e}_p)(\vec{e}_q\vec{e}_m) + 
 (\vec{e}_m\vec{e}_p)(\vec{e}_n\vec{e}_q)]/3$. Calculating 
the energy difference (eq.~\ref{energy-diff-basic}) due to 
the rotation of the atomic moments $i$ and $j$ leads to 
following further terms: 
\begin{eqnarray}
\Delta E_{i,j}^{\mathrm{4-spin}}&=&-K (1+\cos(\theta_i)\cos(\theta_j)-
\cos(\theta_i)-\cos(\theta_j))\nonumber \\
&&-\frac{K}{3} (\sin(\theta_i)\sin(\theta_j)\cos(\phi_i-\phi_j))
\end{eqnarray}
with $K=\sum_{\substack{p \not = q \\ p \not = i,j
\\q \not = i,j}} K_{ijpq}$.

Obviously, one notices that the four-spin 
interactions  with the uncanted environment spins behave for the $ij$ pair like a  bilinear term since only the moments $i$ and $j$ are canted. It is interesting to note that adding this term to 
the Heisenberg model brings an imbalance between the term proportional to the 
$\sin$ function and the one proportional to 
$\cos$. We conclude that this mechanism is behind the observed anisotropy in 
eq.~\ref{last-lloyds}.  
If we restrict ourself to the four-spin interactions only, the difference 
$J_1-J_2$ would be given by $2K/3$ that consequently would lead to a renormalization 
of $J_{ij}$ from $J_2$ to $J_2+\frac{J_2-J_1}{2}$ that we represented as $J_{\mathrm{new}}$ in Fig.~\ref{jij_bulk}. This final result is without any doubt 
subject to modification as soon as higher order terms are 
included in the model Hamiltonian.  The extraction of the exact MEI is thus a rather 
difficult task. As mentioned previously, since the moment of Fe is higher than the one of Ni, the discrepancy between the renormalized $J_{ij}$ and $J_2$ is strongest for Fe (Fig.~\ref{jij_bulk}). 

We exemplify the effect of such corrections by evaluating 
the new Curie temperatures ($T_c$) by  Monte-Carlo simulations. 
The extracted temperatures are not expected to be correct but are 
meant as illustrative examples for the 
effect of renormalizing the MEI. 
A major result shown in Table~\ref{table:Tc} is the large increase of $T_c$ 
with the new values of the MEI for Ni, Co and Fe. 
The difference between the old and new temperature gets stronger when 
increasing the 
magnetic moment of the host. Surprisingly, similar behaviors 
have been obtained by Katsnelson and Lichtenstein\cite{katsnelson} when 
comparing the temperatures 
obtained using the renormalized 
method of Bruno\cite{bruno} with those of the old LKAG method. 
Obviously, the values obtained for Fe are too high and probably, one 
has to include higher order terms in the model Hamiltonian to lower 
$T_c$. The values obtained with only $J_2$ are probably sufficient for Fe due 
to a cancellation of errors that were described by Katsnelson and 
Lichtenstein\cite{katsnelson}. 

\begin{table}[ht!]
\begin{center}
\caption{\label{table:Tc} The Curie temperature (in K) for Ni, Fe and 
Co calculated with the LKAG formalism and with taking into account the 
4-spin interactions.}
\begin{ruledtabular}
\begin{tabular}{lccc}
 $T_c$ (K)         & Exp.  & $J_2$ &  $J_2+\frac{(J_2 - J_1)}{2}$   \\
\hline
  Ni(fcc-LDA)          &631   &  374  &   458 \\
  Co(fcc-GGA)          &1388-1398& 1520 & 1949\\
  Fe(bcc-LDA)          &1045  &  1086 &  2062 \\
  Fe(bcc-GGA)         &      &  1165 &  2791  \\
\end{tabular}
\end{ruledtabular}
\end{center}
\end{table}

By concluding we stress that the LKAG formula for $J_{ij}$ 
describes correctly the MEI for small 
canting angles $\theta$. In this case the spin-dependent $t$-matrices 
$\Delta t$ of Eq.~\ref{tmatrix1} vary linearly in $\theta$, so that 
$\Delta E_{(i,j)}$ is proportional to $J_{ij}~\theta^2$ where $\theta$ is an 
effective canting angle. All higher order interactions like 
$K_{ijkl}$ between 4 or 6 slightly canted spins therefore scale 
as $\theta^4$ or $\theta^6$. As demonstrated, these $J_{ij}$ 
calculated by the LKAG formula include implicitly all multispin interactions 
of the canted $(i,j)$ moments with 
the uncanted environment atoms. It is for these reasons, that the calculated long-wave magnons and the spin stiffness constants agree very well with experiment. However, for larger transversal fluctuations of the moments 
the bilinear interaction $~J_{ij}$ is no longer sufficient, and higher order spin interactions like the four spin interaction and the biquadratic coupling become important and have to be included explicitly in calculating 
$T_c$ and related thermodynamic properties. Since the spin splitting and 
$\Delta t$ scale with the local moments $M$, these multispin interactions 
scale as $M^4$ or higher and are thus more important for systems with large moments. In the paper, we have demonstrated the importantce of four spin interactions in $T_c$-calculations for Fe, Co and Ni based on the LKAG formula for larger canting angles.

We are grateful to J. Kudrnovsky,  
I. Turek and S. Bl\"ugel for fruitful discussions. S. L. wishes to thank the Alexander von Humboldt Foundation
for a Feodor Lynen Fellowship and D. L. Mills for discussions and hospitality.


\begin{thebibliography}{99}

\bibitem{kurz}{Ph.~Kurz, G.~Bihlmayer, K.~Hirai, and S.~Bl\"ugel,
 Phys.~Rev.~Lett {\bf 86}, 
1106 (2001)}

\bibitem{fahnle} {R.~Drautz, M.~Fahnle, Phys. Rev. B {\bf 69}, 104404 (2004).}

\bibitem{molecule}{K.~Park, M.~R.~Pederson, S.~L.~Richardson, N.~Aliaga-Alcalde, G.~Christou,
 Phys. Rev. B, {\bf 68}, 020405(R) (2003)}

\bibitem{alloys}{M.~Lezaic, Ph.~Mavropoulos, S.~Bl\"ugel, Appl.~Phys.~Lett, {\bf 90}, 
82504 (2007)}

\bibitem{alloys2}L. M. Sandratskii, R. Singer, E. Sasioglu, Phys. Rev. B, 
{\bf 76}, 184406 (2007).

\bibitem{surface}{M.~Bode, M.~Heide, K.~von Bergmann, P.~Ferriani, S.~Heinze, 
G.~Bihlmayer, A.~Kubetzka, O.~Pietzsch, S.~Bl\"ugel, R.~Wiesendanger
 Nature {\bf 447}, 190 (2007).}

\bibitem{surface2}L. Udvardi, L. Szunyogh, Phys. Rev. Lett. {\bf 102}, 207204 (2009).

\bibitem{dms1}{M.~Pajda, J.~Kudrnovsky, I.~Turek, V.~Drchal, and P.~Bruno, 
 Phys. Rev. B {\bf 64}, 174402 (2001); G.~Bouzerar {\it et al.}
 Phys.~Rev.~B,{\bf 68}, 81203 (2003).}

\bibitem{dms2}{B.~Belhadji, L.~Bergqvist, R.~Zeller, P.~H.~Dederichs, K.~Sato and H.~Katayama-Yoshida,
 J. Phys.: Condens. Matter. {\bf 19}, 436227 (2007).}

\bibitem{cluster1}S. Lounis, Ph. Mavropoulos, P. H. Dederichs, and S. Bl\"ugel,
Phys. Rev. B {\bf 72}, 224437 (2005); S. Lounis, Ph. Mavropoulos, R. Zeller, P. H. Dederichs, and S. Bl\"ugel Phys. Rev. B {\bf 75}, 174436 (2007);S. Lounis, M. Reif, Ph. Mavropoulos, L. Glaser, P. H. Dederichs, M. Martins, S. Bl\"ugel, W. Wurth, Eur. Phys. Lett. 81, 47004 (2008);
 S. Lounis, P. H. Dederichs, S. Bl\"ugel, Phys. Rev. Lett. {\bf 101}, 107204 (2008).


\bibitem{cluster2}{A. Bergman, L. Nordstr\"om, A. B. Klautau, S. Frota-Pessoa, O. Eriksson, 
~Phys. Rev. B {\bf 73}, 174434 (2006); R. Robles, L. Nordstrom, Phys. Rev. B {\bf 74} 094403 (2006).}

\bibitem{cluster3}O. Sipr, S. Bornemann, J. Minár, S. Polesya, V. Popescu, A. Simunek, and H. Ebert, J. Phys.: Condens. Matter {\bf 19}, 096203 (2007); S. Mankovsky, S. Bornemann, J. Minar, S. Polesya, H. Ebert, J. B. Staunton, A. I. Lichtenstein, 
Phys. Rev. B {\bf 80} 014422 (2009).


\bibitem{cluster4}Ph. Mavropoulos, S. Lounis, S. Bl\"ugel, Phys. Stat. Sol. B {\bf 247}, 1187 (2010); Ph. Mavropoulos, S. Lounis, R. Zeller, S. Bl\"ugel, Appl. Phys. A {\bf 82} 103 (2006).


\bibitem{savrasov2}{X.~Wan, Q.~Yin, S.~Y.~Savrasov, Phys.~Rev.~Lett.~{\bf 97}, 
266403 (2006).}

\bibitem{lichtenstein}{A. I. Liechtenstein, M. I. Katsnelson, V. P. Antropov, 
V. A. Gubanov, J. Magn. Magn. Mat {\bf 67}, 65 (1987).}

\bibitem{heine}{V. Heine, Solid State Phys. {\bf 35}, 1 (1980).}

\bibitem{oswald}{A. Oswald, R. Zeller, P. J. Braspenning and P. H. Dederichs, 
J. Phys. F {\bf 15}, 193 (1985).}

\bibitem{bruno}{P.~Bruno, Phys. Rev. Lett. {\bf 90}, 87205 (2003).}

\bibitem{dederichs}{P.~H.~Dederichs {\it et al.} 
Phys.~Rev.~Lett. {\bf 53}, 2512 (1984).}

\bibitem{antropov2003}{V.~P.~Antropov, J. Magn. Magn. Mat. {\bf 262} L192 (2003); V.~P.~Antropov, 
M.~van~Schilfgaarde, S.~Brink and J.~L.~Xu, 
Journal of Appl.~Phys.~{\bf 99}, 08F507 (2006).}

\bibitem{katsnelson}{M. I. Katsnelson, A. I. Lichtenstein, J. Phys.: Cond. Matter 
{\bf 16}, 7439 (2004).}

\bibitem{callaway} {J.~Callaway, C.~S.~Wang, D.~G.~Laurent, Phys.~Rev.~B 
{\bf 24}, 6491 (1981).}

\bibitem{savrasov}{S.~Y.~Savrasov, Phys.~Rev.~Lett., {\bf 81}, 2570 (1998).}

\bibitem{mills}{A. T. Costa, R. B. Muniz,  D. L. Mills, Phys. Rev. B {\bf 69}, 064413 (2004); 
{\it ibid.} {\bf 70}, 054406 (2004);
{\it ibid.} {\bf 73}, 054426 (2006); A. T. Costa, R. B. Muniz, S. Lounis, A. B. Klautau, D. L. Mills,
{\it ibid.} {\bf 82}, 014428 (2010); A. T. Costa, R. B. Muniz, D. L. Mills,
Phys. Rev. Lett. {\bf 94}, 137203 (2005).}

\bibitem{buczek}P. Buczek, A. Ernst, and L. M. Sandratskii,
Phys. Rev. Lett. {\bf 105} 097205 (2010).

\bibitem{sasioglu}E. Sasioglu, A. Schindlmayr, C. Friedrich, F. Freimuth, and S. Bl\"ugel,
Phys. Rev. B {\bf 81}, 054434 (2010).


\bibitem{staunton}J. B. Staunton, J. Poulter, B. Ginatempo, E. Bruno, and D. D. Johnson,
Phys. Rev. Lett. {\bf 82}, 3340 (1999).

\bibitem{lounis_TDDFT} S. Lounis, A. T. Costa, R. B. Muniz, D. L. Mills, Phys. Rev. Lett. {\bf 105}, 187205 (2010);
{\it ibid.}, ArXiv:1010.1293 (2010); 
A. A. Khajetoorians, S. Lounis, B. Chilian, A. T. Costa, L. Zhou, D. L. Mills, R. Wiesendanger, and J. wiebe, 
ArXiv:1010.1284 (2010).


\bibitem{lloyd}{P. Lloyd, P. V. Smith, Advan. Phys. {\bf 21}, 69 (1972).}

\bibitem{drittler}{B. Drittler, M. Weinert, R. Zeller, P. H. Dederichs, 
Phys. Rev. B {\bf 39}, 930 (1989).}

\bibitem{takahashi}{M. Takahashi, J. Phys. C {\bf 10}, 1289 (1977).}

\bibitem{SKKR}{N. Papanikolaou, R. Zeller, P. H. Dederichs, 
J. Phys.: Condens. Matter {\bf 14}, 2799 (2002).}

\bibitem{LDA}{S.~H.~Vosko, L.~Wilk, M.~Nusair, 
J.~Chem.~Phys.~{\bf 58}, 1200 (1980).}

\bibitem{GGA}{J.~P.~Perdew, Y.~Wang, 
Phys.~Rev.~B {\bf 45}, 13244 (1992).}

\end{thebibliography}
\end{document}